\documentclass[fleqn,10pt]{wlscirep}

\usepackage{color,graphicx}
\usepackage{amsmath,amsfonts,amssymb}
\usepackage{nccmath}

\title{Mitochondrial network complexity emerges from fission/fusion dynamics}

\author[1,5,+,*]{Nahuel Zamponi}
\author[1,+]{Emiliano Zamponi}
\author[2,4]{Sergio A. Cannas}
\author[2,4]{Orlando V. Billoni}
\author[1,4*]{Pablo R. Helguera}
\author[3,4]{Dante R. Chialvo}
\affil[1]{Mitochondrial Research Group, Instituto de Investigaciones M\'edicas Mercedes y Mart\'in Ferreyra y Universidad Nacional de C\'ordoba (INIMEC-CONICET-UNC), Friuli 2434, (5016) C\'ordoba, Argentina.}
\affil[2]{Facultad de Matem\'atica Astronom\'ia F\'isica y Computaci\'on, Universidad Nacional de C\'ordoba, Instituto de F\'isica Enrique Gaviola (IFEG-CONICET), Ciudad Universitaria. (5000) C\'ordoba, Argentina.}
\affil[3]{Center for Complex Systems and Brain Sciences (CEMSC${^3}$), Universidad Nacional de San Martin, Campus Miguelete, 25 de Mayo y Francia (1650), San Mart\'in, Buenos Aires, Argentina.}
\affil[4]{Consejo Nacional de Investigaciones Cient\'ificas y Tecnol\'ogicas (CONICET), Godoy Cruz 2290, (1425) Buenos Aires, Argentina.}
\affil[5]{current address: Department of Medicine, Hematology and Oncology Division, Weill Cornell Medicine, New York, New York 10065, USA.}
\affil[+]{these authors contributed equally to this work}
\affil[*]{zamponi.n@gmail.com, phelguera@immf.uncor.edu}

\begin{abstract}
Mitochondrial networks exhibit a variety of complex behaviors, including coordinated cell-wide oscillations of energy states as well as a phase transition (depolarization) in response to oxidative stress. Since functional and structural properties are often interwinded, here we characterize the structure of mitochondrial networks in mouse embryonic fibroblasts using network tools and percolation theory. Subsequently we perturbed the system either by promoting the fusion of mitochondrial segments or by inducing mitochondrial fission. Quantitative analysis of mitochondrial clusters revealed that the structural parameters of healthy mitochondria lay in  between the extremes of highly fragmented and completely fusioned networks. We confirmed our results by contrasting our emprirical findings with the predictions of a recently described computational model of mitochondrial network emergence based on fission-fusion kinetics. Altogether these results not only offer an objective methodology to parametrize the complexity of this organelle but add weight to the idea that mitochondrial networks behave as critical systems and undergo structural phase transitions.
\end{abstract}

\begin{document}
\flushbottom
\maketitle
\thispagestyle{empty}
Mitochondria arose around two billion years ago from the engulfment of an $\alpha$-proteobacterium by a precursor of the modern eukaryotic cell \cite{evolution}. Subsequent evolution shaped the relation between mitochondria and its host cell, leading to a high degree of specialization of both morphology and function of this organelle. Long known for its role in ATP production, mitochondria also participate in a myriad of essential cellular processes such as apoptosis, calcium buffering and phospholipid synthesis, among others \cite{formfunct}. In addition, mitochondria exhibit complex patterns including oscillatory dynamics, phase transitions and fractality \cite{oscillations, percolation, fractal}.

A typical mitochondria comprise a network of tubule-like structures, with fragments of all sizes (ranging from less than $1\mu m$ to $15 \mu m$ or more) \cite{morpho}.  The current theoretical understanding propose that mitochondrial morphology is maintained by two opposing processes, fission and fusion, which depending on their relative predominance determine the overall connectivity and structural properties of the network \cite{mitdyn}. Although the structure of mitochondrial networks is generally described as complex, a quantitative description of such complexity is still lacking. Moreover, a recent model of mitochondrial dynamics suggested that mitochondrial networks are poissed at the critical point of a phase transition, albeit no connection between theory and phenomenology have been provided yet \cite{Sukhorukov}.

In this work, we present a quantititative description of mitochondrial network structure using tools of network and percolation theory. To do that, we developed a pipeline to extract structural parameters from confocal images. Moreover, we evaluated how a recently published model of mitochondria fitted the data from real networks providing the missing link between theory and experiments.

\section*{Results}
\begin{figure} 
\centering
\includegraphics[width=.9\linewidth]{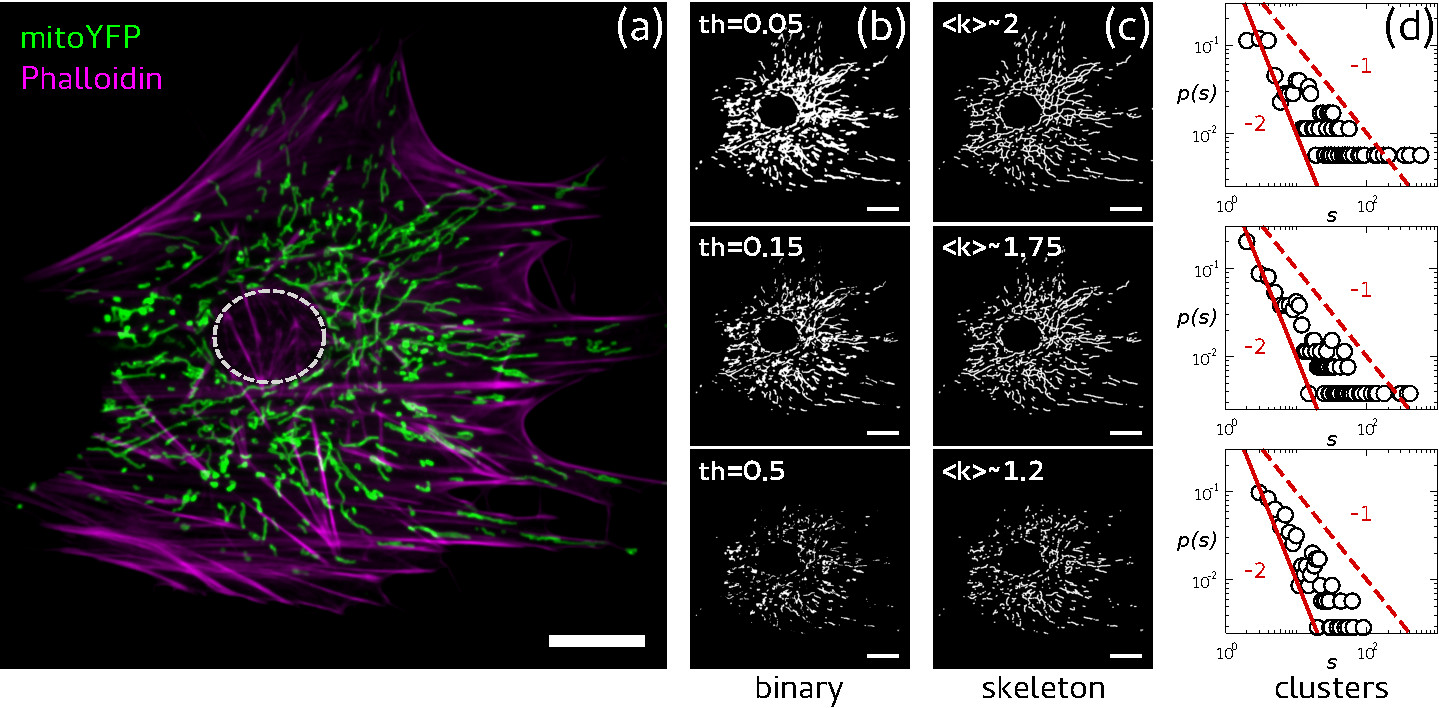}
\caption{\label{fig1} {\it Network structure from mitochondrial images}. {\bf (a)} Confocal image of a MEF in which mitochondria is shown in green and the actin cytoskeleton in magenta. Image processing: {\bf (b)} Grayscale images were converted to binary images by choosing the indicated threshold. {\bf (c)} Detected lines in binary images were reduced to a single pixel width (skeleton). The mean degree $\displaystyle{\langle k \rangle}$ of each skeleton was computed by analysing the nearest neighbors of each pixel occupied. {\bf (d)} Cluster mass distributions $\displaystyle{p(s)}$ were computed by counting the number of pixels that made up each segment within clusters. Red curves represent power laws with exponents -1 (dashed) and -2 (continuous). Scale bars represent 10 $\mu$m.}
\label{fig:1}
\end{figure}

We developed a pipeline to quantify the structural complexity of mitochondrial networks from confocal microscopy images like the one presented in Figure \ref{fig:1}(a), where a mouse embryonic fibroblast (MEF) expressing a mitochondria-targeted yellow fluorescent protein (mYFP) is shown (see Methods). Our pipeline entailed three steps, the first of which was the convertion of a grayscale image into a binary image. To do that, the {\it im2bw} routine from Matlab (Natick, Massachusetts: The MathWorks Inc.) was used. Figure \ref{fig:1}(b) shows examples of binary images obtained from the same grayscale image (Figure \ref{fig:1}(a)) by choosing different threshold values. Second, binarized versions of the image were transformed into skeletons of uniform (one pixel) thickness, like the ones shown in Figure \ref{fig:1}(c), using the routine $bwmorph$ from Matlab. These skeletons were composed of independent clusters of different sizes, each of them made up by either linear or branched segments. Following others \cite{proc_skel_1, proc_skel_2, proc_skel_3}, we hypothesized that such skeletons constituted a good approximation of mitochondrial network structure. Although mitochondrial networks are embeded in the 3-dimensional cellular volume, artefactual branching points in the 2-dimensional reconstructions were negligible (see Supplementary Information). Finally, we extracted two different parameters from skeletons: cluster mass ({\it s}), by counting the number of pixels in each individual cluster, and pixel degree ({\it k}), by counting the nearest neighbors of each pixel (see Methods). Figure \ref{fig:1}(d) shows cluster mass probability distribution functions corresponding to skeletons in Figure \ref{fig:1}(c). We see that, while the connectivity decreases when the threshold increases (as expected due to a larger network fragmentation), $p(s)$ exhibit a power law behavior that seems to be robust against threshold variations.

\begin{figure} 
\centering
\includegraphics[width=0.7\linewidth]{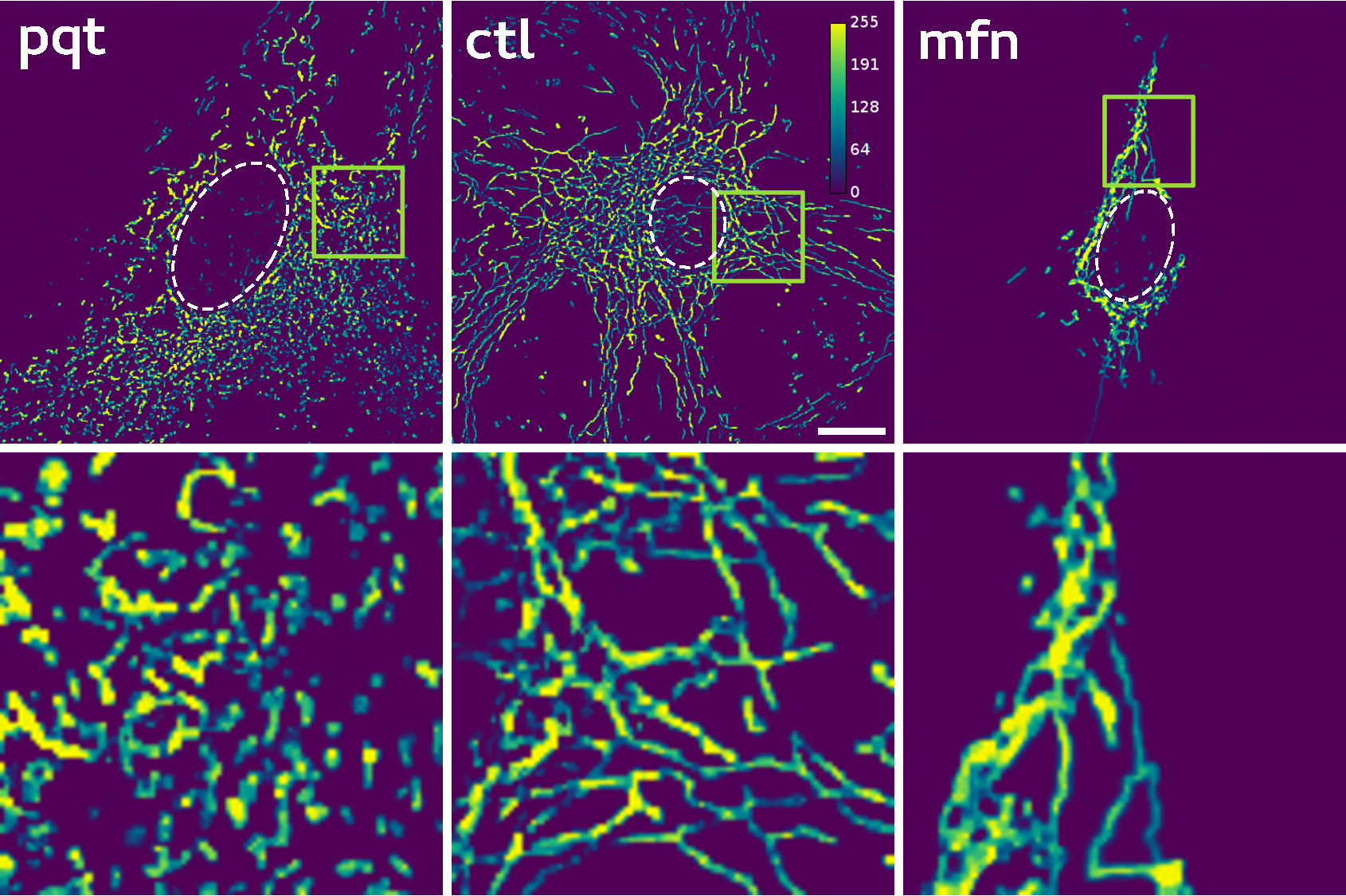}
\caption{\label{fig2} {\it Typical examples of mitochondrial network images obtained under different conditions}. Confocal images of MEFs transfected with mYFP under control conditions ({\it ctl}), paraquat treatment ({\it pqt}) or mitofusin 1 over-expression ({\it mfn}). Pixel intensity is depicted using a pseudocolor (calibration bar). Insets highlight the effect asserted by each treatment on mitochondrial structure. Scale bars represent 10 $\mu$m.}
\label{fig:2}
\end{figure}

The long-tail behaviour displayed by $\displaystyle{p(s)}$ is in accordance with recent work suggesting that mitochondrial networks operate near a percolation phase transition \cite{Sukhorukov}. In order to test the existence of a structural transition, we perturbed the structure of mitochondrial networks in opposite directions, either by increasing mitochondrial fission using paraquat ({\it pqt}) \cite{pqt_1, pqt_2, pqt_3, pqt_4}, or by promoting mitochondrial fusion by mitofusin 1 ({\it mfn}) over-expression \cite{mfn_1, mfn_2, mfn_3, mfn_4} (see Methods). Examples of the morphological changes observed after applying the aforementioned treatments are presented in Figure \ref{fig:2}. A rapid qualitative inspection revealed that, compared to control ({\it ctl}) networks, {\it mfn} networks appeared as elongated interconnected strings, while in the case of the {\it pqt} networks, mitochondria seemed as independent small fragments. In fact, this qualitative visual inspection is the approach used routinely to evaluate mitochondrial morphology and status \cite{qual1, qual2, qual3, qual4, qual5, qual6}, counting the relative ratios of cells exhibiting a particular  mitochondrial phenotype, different from the one found in control cells.

\subsection*{Quantification of mitochondrial network properties}
\begin{figure}
\centering
\includegraphics[width=.9\linewidth]{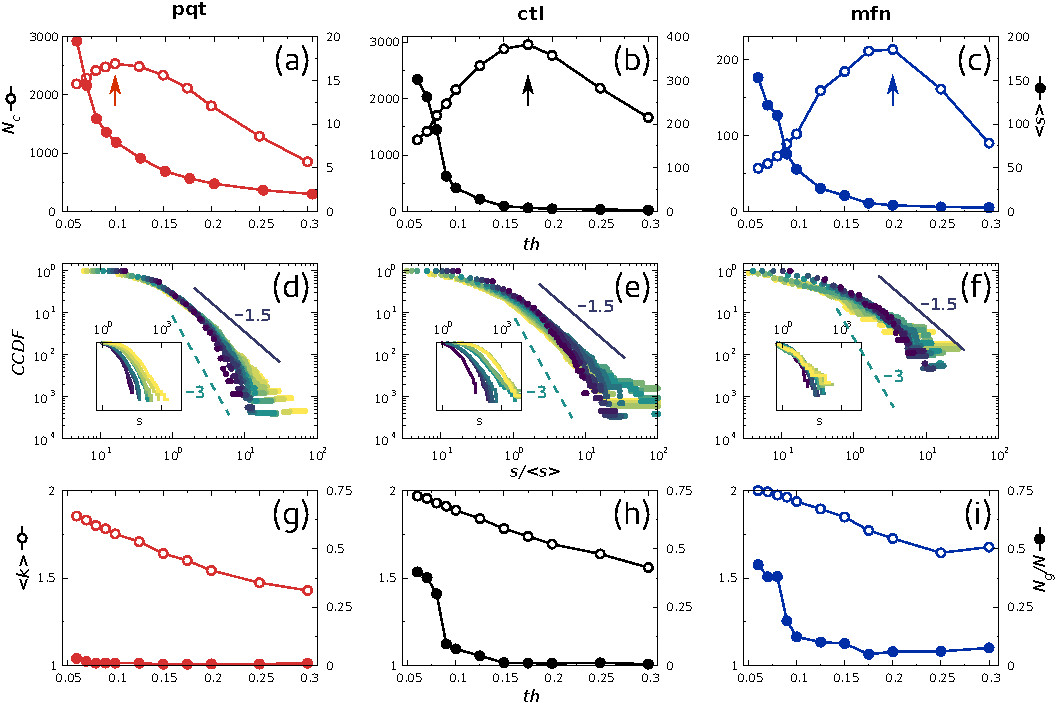}
\caption{\label{fig3} {\it Network parameters computed from single images}. Center column corresponds to {\it ctl}, left column to {\it pqt} and right column to {\it mfn}. Top row panels ({\bf (a)}, {\bf (b)} and {\bf (c)}) show the number of clusters ($N_{c}$, empty symbols, left labels) and the mean cluster mass ($\langle s \rangle$, filled symbols, right labels) as a function of the binarization threshold ($th$). Arrows point to $th$ values at which the maximum number of clusters appear in each case. Middle row panels ({\bf (d)}, {\bf (e)} and {\bf (f)}) show the cumulative distribution function of cluster mass as a function of {\it th} (colors denote different threshold values). Note that by normalizing each distribution using the mean cluster mass $\langle s \rangle$, all distributions collapse approximately to the same function. Power laws with exponents $-1.5$ and $-3$ are shown for reference purposes. Insets depict unnormalized distributions. Bottom row panels ({\bf (g)}, {\bf (h)} and {\bf (i)}) depict the mean degree ($\langle k \rangle$, empty symbols, left labels) and the normalized size of the largest cluster ($N_{g}/N$, filled symbols, right labels) of the networks as a function of {\it th}.}
\label{fig:3}
\end{figure}

Although many methods have been proposed to extract networks from mitochondrial images, virtually all of them made use of an arbitrary thresholding step \cite{proc_skel_1, proc_skel_2, proc_skel_3, th_1, th_2, th_3}. However, as illustrated in Figure \ref{fig:1}, the application of different intensity thresholds ({\it th}) resulted in different skeletons and hence different network architectures. We reasoned that a way to avoid this arbitrariness would be to compare the behavior of network parameters as a function of the threshold. As a proof of concept, we picked one network for each condition and computed basic structural parameters. The results obtained are shown in Figure \ref{fig:3}, where each column corresponds to one of the treatments: {\it ctl} in the middle, {\it mfn} on the right and {\it pqt} on the left. Similar results were obtained using other images (data not shown).

As depicted in Figure \ref{fig:3}(a), (b) and (c), we first looked at the number of clusters $N_{c}$ and the average cluster mass $\langle s \rangle$ of the networks \cite{IntroPerc}. $N_{c}$ exhibited a non monotonic behavior in all cases, with a peak in the interval $0.1 \leq th^* \leq 0.2$ (Figure \ref{fig:3}(a), (b) and (c), empty circles). This is expected, regardless of the nature or origin of the image, anytime a section is made cutting a rough landscape \cite{Nc_avS, modurbgropat}. Interestingly, the peaks of {\it pqt} and {\it mfn} networks shifted to left and to right, respectively, from the peak of the {\it ctl} network, suggesting that the relative position of the peak could constitute a robust readout of network connectivity. Moreover, we enviosioned that the position of the peak in $N_{c}$ could be usefull as a less arbitrary criterion when the selection of a single threshold is needed.\\
On the other hand, $\langle s \rangle$ changed monotonically with the threshold in all cases (Figure \ref{fig:3}(a), (b) and (c), solid circles). Regardless of the similar behavior, $\langle s \rangle$ values of the {\it pqt} network were much smaller than that of other networks, in accordance with the effect asserted by the compound on mitochondria. In the case of the {\it mfn} network, although its $\langle s \rangle$ values were within the same order of magnitude of those of the {\it ctl} network, it should be noted that mitofusin 1 over-expression reduced overall network size (see Figure \ref{fig:2}), making them proportionally bigger.

Long-tail distributions in Figure \ref{fig:1}(d) indicated that mitochondrial networks exhibit scale free properties, such as the coexistence of numerous small fragments with few massive clusters.
To extract more precise information from mass distributions, we plotted the complementary cummulative distribution function (CCDF)
\begin{ceqn}
\begin{align}
P(s) = \sum_{s_{i} \geq s} p(s_{i})
\end{align}
\end{ceqn}
where $p(s_{i})$ is the probability of finding a cluster of mass $i$ in the network, which enhance the statistical significance of the high mass region \cite{ccdf}. Figure \ref{fig:3}(d), (e) and (f) show the CCDFs of cluster mass as a function of a range of threshold values. CCDFs describe how often the cluster mass is above a particular value $s$ and we used them here to characterize the effect of the treatment on the giant cluster of the network. In addition, we found that the average cluster mass $\langle s \rangle$ can be used to normalize mass densities, eventually leading to a unique average representative distribution for any threshold value that followed approximately a power-law of the form
\begin{ceqn}
\begin{align}
P(s) \sim s^{-\gamma + 1}
\end{align}
\end{ceqn}
with $ 0.5 \leq \gamma \leq 2$. This result allowed us to consider the cluster mass distribution as roughly independent of the threshold value, which is a desirable property in any descriptive parameter used to quantify the structure of mitochondrial networks. More importantly, Figure \ref{fig:3}(d), (e) and (f) suggest that perturbations on the fission/fusion balance altered mass frequencies and modified the mass of the giant cluster, causing a shift in the slope of the CCDF. 

Figure \ref{fig:3}(g), (h) and (i) describe two additional topological quantities inspired in network theory, namely the average degree $\langle k \rangle$ (empty dots) and the normalized size of the giant cluster $N_g/N$ (solid dots) \cite{Erdos}. Although $\langle k \rangle$ decreases monotonically in all cases, the relation $\langle k \rangle _{pqt} < \langle k \rangle _{ctl} < \langle k \rangle _{mfn}$ holds for every particular threshold value, indicating once again that perturbing the fission/fusion balance drove network connectivity to the extremes of disconnected fragments or fully connected cluster. 
Finally, when we normalized the mass of the giant clusters by computing $N_g/N$ we found that, despite the effects that the treatments asserted on network size, the {\it ctl} network displayed an intermediate behaviour compared to {\it pqt} and {\it mfn} networks.

\begin{figure}
\centering
\includegraphics[width=.8\linewidth]{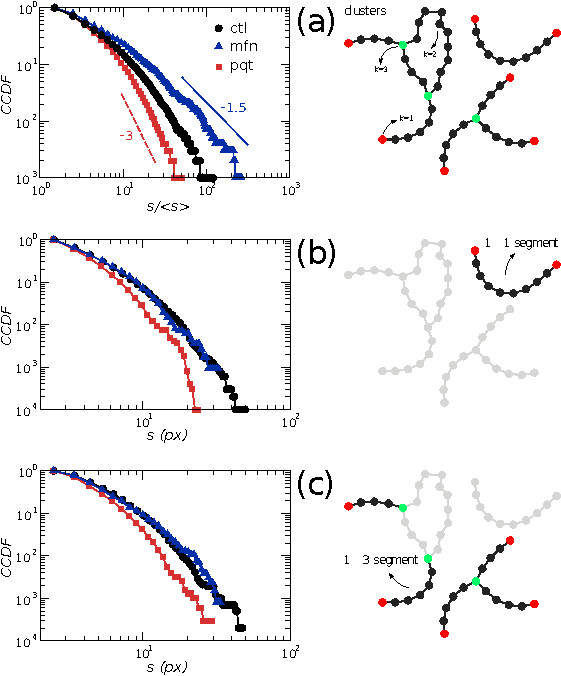} 
\caption{\label{fig4} {\it Changes in mass distributions upon fission/fusion balance perturbation}. {\bf (a)} Cluster mass distributions followed a truncated power law. Shifts in the behavior upon treatment obeyed the relation $ \gamma_{mfn} < \gamma_{ctl} < \gamma_{pqt}$, where $\gamma$ is the exponent of the CCDF. Power laws with exponents $-1.5$ and $-3$ are shown for reference purposes only. Mass distribution of open {\bf (b)} and branched {\bf (c)} segments showed exponential decays. Illustrations at the right side of each panel depict the network elements measured.}
\label{fig:4}
\end{figure}

The results presented in Figure \ref{fig:3} suggested that mitochondrial network structure changed in response to treatments in a rather predictable manner, with {\it ctl} networks always displaying an intermediate behavior. To gain insight into how mitochondrial structure responded to perturbations we recomputed the CCDFs from Figure \ref{fig:3}(d),(e) and (f) using all the available data, by selecting a threshold value of $0.15$. Similar results were obtained using thresholds in the range [0.1-0.2] (data not shown).  Schemes in Figure \ref{fig:4}(a), (b) and (c) define the elements measured in every network, namely clusters and segments, where 1-1 means a segment that connects two nodes with $k=1$ and 1-3 refers to those segments connecting a node with $k=3$ and a node with $k=1$.
Figure \ref{fig:4}(a) shows that modifications of the fission/fusion balance changed the exponents of the power-law relation in the cluster mass distribution, adding support to our previous findings. Interestingly, mass distributions of 1-1 and 1-3 segments behaved similarly accross treatments, suggesting that differences between networks could not be attributed to changes in the mass of linear segments.

\subsection*{Fission/fusion balance is linked to mitochondrial network complexity}
In the previous section we have described a straightforward methodology and a set of structural parameters that allowed us to characterize basic features of mitochondrial networks. To gain a deeper understanding of the relation between mitochondrial dynamics and network status we computed additional parameters in order to quantify the complexity of the different networks.

\begin{figure} 
\centering
\includegraphics[width=.9\linewidth]{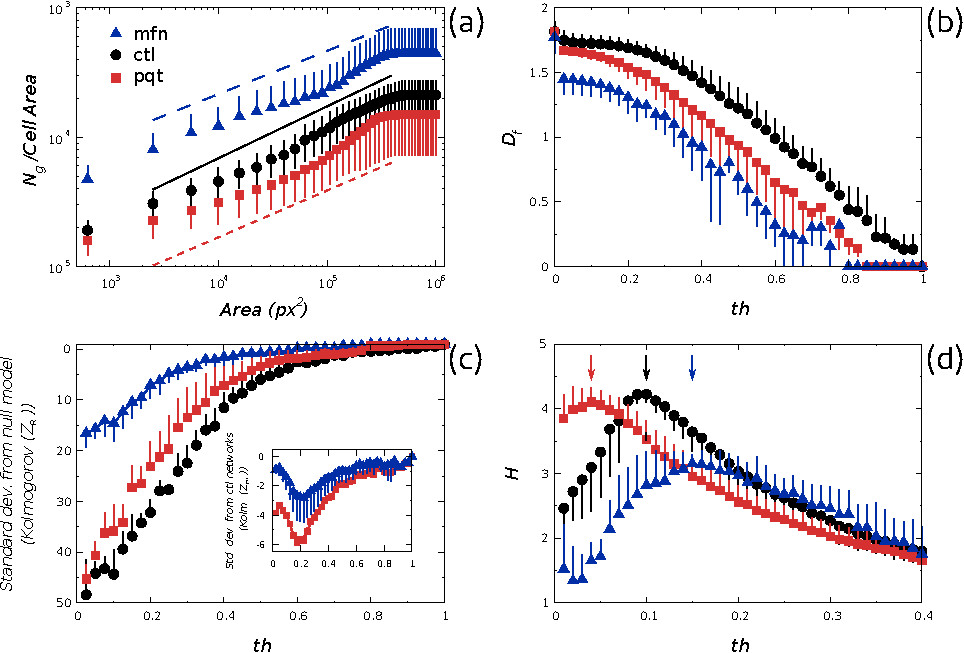}
\caption{{\it Changes in mitochondrial network complexity}. {\bf (a)} Scaling of $N_g/N$ with the area used to sample the image. {\bf (b)} Changes in fractal dimension upon paraquat treatment and mitofusin 1 over-expression. {\bf (c)} $Z$-scores expressing the deviation of the observed structure from the random configuration in Kolmogorov complexity units. Inset shows the deviation of {\it pqt} and {\it mfn} network configurations from the {\it ctl} configuration in Kolmogorov complexity units. {\bf (d)} Entropy of cluster mass distributions. Perturbations of the fission/fusion balance caused a shift in the critical threshold value $th^{*}$ (arrows).}
\label{fig:5}
\end{figure}

We investigated the scaling properties of mitochondrial networks by performing a finite size scaling analysis \cite{scaling1, scaling2}. Results are summarized in Figure \ref{fig:5}(a) where the normalized mass of the giant cluster as a function of the area used to sample the image is shown. Fitted lines showed that scaling behavior of {\it pqt} and {\it mfn} networks slightly deviated from the one displayed by {\it ctl} networks, suggesting that structural alterations caused by the treatments affected network scaling.

Next, given the functional relevance of the spatial distribution of mitochondria \cite{ERmit, kif, mtDNA, interactome}, we decided to quantify the fractal dimension $D_{f}$ of the networks \cite{Df1, Df2}. As can be seen in Figure \ref{fig:5}(b), even though treatments had opposite effects on mitochondria, they both decreased the network $D_{f}$, suggesting once again that alterations in the fission/fusion balance impacted on the space-filling properties of mitochondria. 

These two observations prompted us to hypothesize that alterations in the fission/fusion balance tended to reduce network complexity. To test this hypothesis we computed the normalized Kolmogorov complexity \cite{Kolmogorov}
\begin{ceqn}
\begin{align}
Z_{R}=\frac{K-\mu_{R}}{\sigma_{R}^2}
\end{align}
\end{ceqn}
where $K$ is the Kolmogorov complexity of the network, $\mu_{R}$ is the mean Kolmogorov complexity from 1000 randomized versions of the original skeleton and  $\sigma_{R}^2$ is the standard deviation from the randomized versions of the original skeleton. As shown in Figure \ref{fig:5}(c), the distance between the observed network structure and the random configuration, in Kolmogorov complexity units, was maximal for {\it ctl} networks, strongly supporting the idea that perturbations to mitochondrial fission or fusion lowered network complexity.

Finally, we considered the possibility that changes in network parameters could be explained as shifts in the percolation regime of the networks, taking into account recent suggestions that healty mitochondrial networks are in a critical regime, characterized by the maximal heterogeneity in sizes of the network subcomponents \cite{Sukhorukov}. Specifically, we reasoned that the percolation transition threshold $th^{*}$ of the images could indicate the stage of the percolation process of the underlying networks, {\it i.e.}, {\it subcritical}, {\it critical} or {\it supercritical} regime \cite{CritPhen}. To test this, we computed the Shannon's entropy of cluster masses \cite{shannon1, shannon2}
\begin{ceqn}
\begin{align}
H_{j} = -\sum_{\substack{ \forall i}} p_{i,j} log(p_{i,j})
\end{align}
\end{ceqn}
where $H_{j}$ is the entropy at threshold $j$ and $p_{i,j} = \frac{\sum M_{i,j}}{M_{j}}$ is the fraction of the total mass of clusters of mass $i$ at threshold $j$ over the total mass of the network at threshold $j$. Plotting $H_{j}$ as a function of the threshold (Figure \ref{fig:5}(d)) we observed that each network type was characterized by a particular $th^{*}$ value. Moreover, the following relation was found $th_{pqt}^* \leq th_{ctl}^* \leq th_{mfn}^*$, indicating that configurations found in {\it pqt} and {\it mfn} networks corresponded to {\it subcritical} and {\it supercritical} regimes, respectively.

In summary, our results showed that perturbations to fission/fusion kinetics gave rise to changes in connectivity patterns, altering the scaling properties and the percolation regime of mitochondrial networks.
\subsection*{Are mitochondrial networks poised at criticality?}
The structural and topological changes described in the previous paragraphs were consistent with the idea that mitochondrial networks undergo a percolation transition \cite{IntroPerc, PercTran1, PercTran2}. This view propose that a steady state mitochondrial network requires a proper balance of two opposing tendencies, one toward fusing segments and the other to favor fragmentation. 

In an attempt to further test these interpretations we contrasted our experimental results with the predictions of a recently published mathematical model \cite{Sukhorukov} that contains explicit variables for relative fission/fusion rates $c_1$ and $c_2$ (see Methods). 
In order to proceed, a bootstrapping approach was required to extract model parameters from the experimental data. First, the order parameter was defined as the ratio of the largest cluster $\langle N_g\rangle$ over the network size $N$. Then, for a fixed value of $c_1$ both the order parameter ($\langle N_g\rangle/N$) and the average degree $\langle k \rangle$ as a function of $c_2$ were numerically computed. Next, for every value of $c_1$, $\langle N_g\rangle/N$ was plotted parametrically by varying $c_2$, as a function of $\langle k \rangle$ (See Methods).

\begin{figure} 
\centering
\includegraphics[width=.9\linewidth]{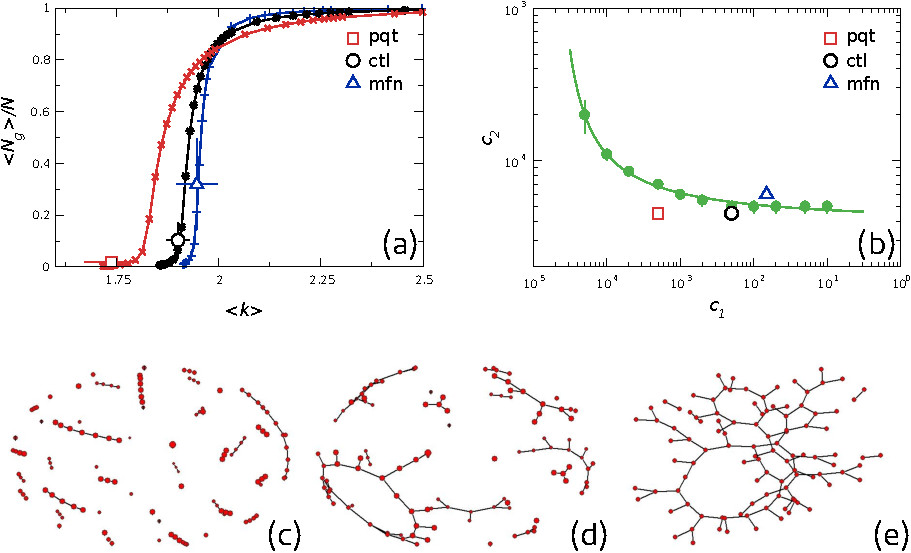} 
\caption{\label{fig5} {\it Comparison of the present experimental results with those of Sukhorukov' {\it et al.} model}. (a) Model parameters extracted iteratively from the experimental data. Open symbols and error bars correspond to means and standard deviations of all cells for the {\it ctl}, {\it pqt} and {\it mfn} groups (for a binarization threshold = 0.15). Each point of the extracted curves corresponds to a unique pair of values $(c_1,c_2)$.
(b) Phase diagram of the model in the $(c_1,c_2)$ space. Filled symbols and the continuous line correspond to the location of the phase transition. The three open symbols (labeled {\it pqt}, {\it ctl}, {\it mfn}) correspond to the parameter values extracted from the experimental data as shown in (a). Panels (c), (d) and (e) show a graphical representation of the typical networks simulated using the three derived $(c_1,c_2)$ values, corresponding to {\it pqt}, {\it clt} and {\it mfn} networks, respectively.}
\label{fig:6}
\end{figure}

Continuous lines in Figure \ref{fig:6}(a) are an example of those curves. It could be observed that, at least for the parameters region of interest, each point of the curves corresponded to a unique pair of values $(c_1,c_2)$ (see Supplementary Information). This allowed us to roughly associate model parameters values to the experimental data. The symbols in the figure correspond to an average of $\langle N_g\rangle/N$ and $\langle k \rangle$ over different cell groups. The phase diagram in the $(c_1,c_2)$ space is illustrated in Figure \ref{fig:6}(b). The filled symbols (with error bars) correspond to the maxima in the mean cluster size $\langle s \rangle$ (see Supplementary Information). The continuous line is a non linear fitting to the points and represents a reference to the eye for the location of the phase transition. The three symbols correspond to the parameter values extracted from the experimental data in the three conditions shown in {\bf (a)}. The graphs on panels {\bf (c)}, {\bf (d)} and {\bf (e)} show examples of the typical networks constructed with the model using the three derived empirical values (i.e., the points in {\bf (b)} labeled {\it pqt}, {\it ctl}, {\it mfn} respectively). From Figure \ref{fig:6} it can be concluded that {\it ctl} networks were in the vicinity of a percolation phase transition, while {\it pqt} and {\it mfn} network configurations corresponded to {\it subcritical} and {\it supercritical} regimes, respectively.

\section*{Discussion}

Mitochondria organize as complex networks that display temporal and spatial coordination, pressumably by operating close to the edge of dynamic instability \cite{percolation}. Following these ideas, in this work we explored the topological properties of mitochondria and we tested the hypothesis that mitochondrial networks are organized near the percolation transition \cite{CritPhen}.

The main findings of the present study can be summarized in three aspects. First, we proposed a straightforward approach to quantify structural properties of mitochondrial networks. Applying the proposed method to control ({\it ctl}) and perturbed ({\it pqt} and {\it mfn}) networks, we identified regularities in the variations of connectivity patterns as well as changes in the behavior of the mass distribution of mitochondrial clusters. Moreover, data from additional experimental manipulations supported the conclusions obtained with paraquat treatment and mitofusin 1 over-expression (see Supplementary Information).

Second, by computing additional parameters we determined that the configuration found in control mitochondria is optimal. Specifically, we observed that promoting fission or fusion lowered the fractal dimension of the networks \cite{Df1}, reducing the space-filling capacity of mitochondria. Additionally, we found that promoting either fission or fusion lowered the Kolmogorov complexity of mitochondrial networks \cite{Kolmogorov}, increasing the randomness of their configuration.

Finally, we contrasted our empirical data with a recently published mathematical model \cite{Sukhorukov} finding that balanced fission/fusion dynamics lead to a network capable of a phase transition. Specifically, the empirical network configurations were found at parameter regions predicted by the model, allowing us to classify network configurations in three different regimes, {\it subcritical} ({\it pqt}), {\it critical} ({\it ctl}) and {\it supercritical} ({\it mfn}), and adding support to the idea that, under normal physiological conditions, mitochondrial networks are poissed near a percolation transition point.

\section*{Methods}
\subsection*{Cell culture}
Mouse Embryonic Fibroblasts (MEFs) were obtained as described by Xu {\it et al}\cite{mefs}. Briefly, 13.5 days old mouse embryos were extracted from the mother uterus, rinsed with PBS and placed on a petri dish. The head and red organs were discarded and the remaining body was rinsed again with PBS and placed on a new petri dish. Using shaving blades, the tissue was chopped into little pieces and trypsinized for 15 sec at 37 $^\circ$C (PBS 10\% Trypsin). Trypsin reaction was quenched with serum-containing media and the whole mixture was centrifugued for 5 min at 2000 rpm. The supernatant was discarded and the pellet resuspended in DMEM with 10\% SFB, 1\% GlutaMax and 1\% Non-essential amino acids. Cell pellets from 4 embryos were seeded on 175 cm$^{2}$ culture bottles and were allowed to grow for 48 h. C57BL6 mice were obtained from the Animal Facility of the Mercedes and Mart{\'i}n Ferreyra Medical Research Institute and National University of C{\'o}rdoba (INIMEC-CONICET-UNC). All experimental protocols were approved by the Institutional Council of Animal Care (CICUAL-INIMEC-CONICET). All methods were carried out in accordance with the approved guidelines.

\subsection*{Mitochondrial network morphology manipulation and cell imaging}
Mitochondria was visualized by lentiviral infection. Lentiviruses were produced as described by Baloh {\it et al.} \cite{lentivirus}. Briefly, human embryonic kidney (HEK) 293T cells were plated onto six-well plates and transfected with a polymerase-coding vector (REV), a packaging vector (8.71), an envelope vector (VSVG) and a shuttle vector encoding the mitochondrial-targeted yellow fluorescent protein (mitoYFP) using Lipofectamine 2000 (Invitrogen) reagent. Media was changed at 12 h and collected at 48 and 72 h, pooled, and applied directly to MEFs cultures.

Manipulation of mitochondrial morphology was accomplished in two ways: 1) to induce mitochondrial fragmentation, mitoYFP-expressing MEF cultures were treated with $200 \mu M$ paraquat (Sigma-Aldrich) for 24 hours, and 2) to promote mitochondrial fusion, cultures were transfected (Lipofectamine 2000, Invitrogen) with a plasmid encoding the sequence of the human mitofusin 1 gene (MFN1, Addgene). Once ready, cells were fixed with 4\% PFA in 4\% sucrose-containing PBS. F-actin staining was performed on fixed cells using Alexa 546-phaloidin (Molecular Probes), following manufacturer's protocol. In all cases, image acquisition was achieved using an Olympus IX81 inverted microscope equipped with a Disk Spinning Unit (DSU), epifluorescence illumination (150 W Xenon Lamp) and a microprocessor. MEFs were imaged using a 60x oil immersion objective, an ORCA AG (Hamamatsu) CCD camera and OSIS software.

\subsection*{Image analysis}
All routines used for image processing and analysis were written in Matlab (The MathWorks, Natick, MA). As explained above, mitochondrial structures were extracted from micrographs of MEFs where mitochondria were fluorescently tagged. Individual 8-bit images were converted to binary (i.e., black \& white) for different threshold values of intensity using the Matlab routine $im2bw$. For each threshold value (range 0-1) the skeleton (i.e., the image reduced to a trace of one- pixel thickness) was extracted using the Matlab routine $bmorph$. Subsequently, clusters were extracted using the Matlab routine $bwlabel$. The algorithm define a cluster as those pixels connected with at least one of the eight nearest neighbors. The degree was computed using a numerical routine that inspects the nearest neighbors of each pixel and decides if the site corresponds to a node of degree $k=1$, $k=2$ or $k=3$ \cite{NEFI}. $Ng/N$ scaling was computed by averaging the size of the biggest cluster in sections of the image of increasing area. It was determined that the reconstructed network topology was not affected by the issue of potential artifacts due to the 2D projection of the mitochondrial 3D structure (see Supplementary Information). The complementary cumulative distribution function is defined as
\begin{ceqn}
\begin{align}
CCDF = Pr(X \ \geq \ x)
\end{align}
\end{ceqn}
and was computed by calculating the fraction of clusters of mass higher than $x$, were $x$ takes the value of all possible cluster masses. Fractal dimension ($D_{f}$), Kolmogorov complexity and Entropy ($H$) were computed using the Matlab routines $boxcount$, $kolmogorov$ and $entropy$, respectively.

\subsection*{Sukhorukov \textit{et al.} Model}
Numerical simulations were conducted using the model described in Sukhorukov \textit{et al.} \cite{Sukhorukov}.
Briefly, the network structure emerges as the result of two fusion and two fission reactions between the tips of a set of $L$ dimers. In the model, a dimer tip can connect (disconnect) to other dimer tips forming a network node, but at most three tips can be merged. In this way, the degree $k$ of the nodes can take only the values $k=1$ (isolated tip), $k=2$ (two merged tips) and $k=3$ (three merged tips); only two fusion processes are allowed: tip-to-tip (two nodes of degree $k=1$ merge into a node of degree $k=2$) and tip-to-side (a node of degree $k=1$ and a node of degree $k=2$ merge into a node of degree $k=3$). To each fusion process there is an associated inverse (fission) one. The bias to each process can be written as rates of either fusion or fission\cite{modeloref1} represented as reaction processes on nodes $X_k$:
\begin{ceqn}
\begin{align}
2X{_1}_{\overleftarrow{b_1}}^{\overrightarrow{a_1}}X{_2}
\end{align}
\end{ceqn}
\begin{ceqn}
\begin{align}
X{_1} +X{_2}_{\overleftarrow{b_2}}^{\overrightarrow{a_2}}\textrm{X}_3
\end{align}
\end{ceqn}
where $a_1$ ($b_1$) is the reaction rate for tip-to-tip fusion (fission) process and $a_2$ ($b_2$) is the reaction rate for tip-to-side fusion (fission) process. The model can then be implemented as an agent based stochastic dynamics between a set of $L$ reactant objects (dimers), submitted to the above described fusion and fission processes. The dynamics is simulated using Gillespie\cite{Gillespie} algorithm. Nodes participating in a particular event are chosen with equal probability within a list of the nodes with the same degree. Following Ref.\cite{Sukhorukov} we assumed $b_2=(3/2)b_1$ and varied the relative rates $c_i=a_i/b_i$.

Sukhorukov \textit{et al.} described in detail the steady state of the dynamics as a function of changes in $c_1$ and $c_2$ \cite{Sukhorukov}. The system admits a plethora of network configurations in parameter space, including fragmented or hyperfused networks resulting from extreme values of fission and fusion activities as well as networks resembling those seen in healthy cells at intermediate values. In passing, notice that more recently this model was reformulated \cite{Sukhorukov2} to include information on the microtubule cytoskeleton.

In the present work we performed simulations for $L=15000$, roughly the estimated value for the average number of edges in the control cell images. In every simulation we run the algorithm $3L$ times after which we measured different quantities. This ensured that the distribution of nodes with degree $k$ became stationary. For every set of values of $(c_1,c_2)$ the procedure was repeated 100 times for different sequences of random numbers and the different quantities were averaged over this sample. The different quantities measured were: the average degree $\langle k \rangle$, the average fraction of nodes in the largest cluster $\langle N_g/N \rangle$ and the average cluster size excluding the largest cluster $\langle s \rangle$ (see Supplementary Information), where the averages were taken both over all the nodes in the network and over different runs.
\\
\\
{\bf Data availability}. The data that supports the findings presented in this study is available upon request.

\section*{Acknowledgements}
This work was partially supported by CONICET (Argentina) through Grants PIP 112-201101-00213, PIP 2015-01-00954 and PICT 2013-3142, and by SECyT (Universidad Nacional de Córdoba, Argentina).
DRC thanks the support of Universidad Nacional de San Martín, Argentina.

\section*{Author contributions statement}
NZ, EZ, PH and DC conceived the experiments; NZ and EZ conducted the experiments; NZ, EZ, SC, OB and DC analyzed the data. All authors reviewed the manuscript.

\section*{Additional information}

{\bf Supplementary information} accompanies this paper at http://www.nature.com/srep
\\
\\
{\bf Competing financial interests:} The authors declare no competing financial interests.\\
\\
{\bf How to cite this article:}

\end{document}